\documentclass[manuscript]{aastex}

\shorttitle{Extreme solar particle events}
\shortauthors{Usoskin et al.}
\usepackage{natbib}

\begin{document}


\title{Occurrence of extreme solar particle events: Assessment from historical proxy data}


\author{Ilya G. Usoskin}
\affil{Sodankyl\"a Geophysical Observatory (Oulu unit) and Dept. of Physical Sciences, University of Oulu, Finland}
\email{ilya.usoskin@oulu.fi}

\and

\author{Gennady A. Kovaltsov}
\affil{Ioffe Physical-Technical Institute of RAS, 194021 St.Petersburg, Russia}




\begin{abstract}
The probability of occurrence of extreme solar particle events (SPEs) with the fluence of
 ($>30$ MeV) protons $F_{30}\ge 10^{10}$ cm$^{-2}$ is evaluated based on data of cosmogenic
 isotopes $^{14}$C and $^{10}$Be in terrestrial archives centennial-millennial time scales.
Four potential candidates with $F_{30}=(1\div 1.5)\cdot 10^{10}$ cm$^{-2}$ and no events with $F_{30}>2\cdot 10^{10}$ cm$^{-2}$
 are identified since 1400 AD in the annually resolved $^{10}$Be data.
A strong SPE related to the Carrington flare of 1859 AD is not supported by the data.
For the last 11400 years, 19 SPE candidates with $F_{30}=(1\div3)\cdot 10^{10}$ cm$^{-2}$ are found
 and clearly no event with $F_{30}>5\cdot 10^{10}$ cm$^{-2}$ (50-fold the SPE of 23-Feb-1956) occurring.
This values serve as an observational upper limit for the strength of SPE on the time scale of tens of millennia.
Two events, ca. 780 and 1460 AD, appear in different data series making them strong candidates to
 extreme SPEs.
We built a distribution of the occurrence probability of extreme SPEs,
 providing a new strict observational constraint.
Practical limits can be set as $F_{30}\approx 1$, 2$\div$3, and 5 $10^{10}$ cm$^{-2}$ for the occurrence probability
 $\approx 10^{-2}$, $10^{-3}$ and $10^{-4}$ year$^{-1}$, respectively.
Because of uncertainties, our results should be interpreted as a conservative upper limit of the SPE occurrence
 near Earth.
The mean SEP flux is evaluated as $\approx 40$ (cm$^2$ sec)$^{-1}$ in agreement with estimates from the lunar rocks.
On average, extreme SPEs contribute about 10\% to the total SEP fluence.
\end{abstract}





\section{Introduction}

Sporadic energy releases on the Sun can accelarate solar energetic particles
 (SEP) in the corona and interplanetary medium.
Such phenomena often lead to solar particle events (SPEs) observed at Earth, that is an
 important factor of solar-terrestrial relation, and specifically Space Weather \citep{mewaldt06,vainio09}.
A typical quantity for a SPE is the fluence of SEP with energy above 30 MeV, $F_{30}$.
We have sufficient knowledge of SPEs over the space era since mid-1950s \citep{smart06}, with
 only several events with $F_{30}=(1\div 10)\cdot 10^{9}$ cm$^{-2}$ and hundreds of weaker SPEs observed.
However, it is important to know, both for purely theoretical aspects of solar/stellar physics and
 for technical applications, the statistics of extreme SPEs with $F_{30}>10^{10}$ cm$^{-2}$.
Such a study is possible only using indirect proxy data.
Some estimates have been obtained from the measurements of cosmogenic isotopes lunar rocks \citep{nishiizumi09}
 but it gives only the average flux of SEP over very long scale without extracting individual SPEs.
A list of potential SPEs over the last 500 years was proposed based on nitrate records in polar ice \citep{mccracken01},
 but this result is heavily debated \citep{wolff08,wolff12}.
Cosmogenic nuclide $^{14}$C and $^{10}$Be data measured in terrestrial archives may provide
 information on SPE in the past \citep{lingenfelter80,usoskin_GRL_SCR06,webber07} but this possibility was not
 fully explored earlier.
Accordingly, the probability of extreme SPEs remained grossly uncertain \citep{hudson10}.

Here we establish a solid observational constraint on the distribution of extreme SPEs using
 presently available datasets of cosmogenic isotopes ($^{14}$C and $^{10}$Be) measured in
 terrestrial archives with sufficient time resolution and quality, and modern models of their production in the atmosphere.
We note that the presented resulted is based on terrestrial data and may not well represent
 the occurrence of solar events, whose geo-efficiency is also affected by the relative Sun-Earth attitude.

\section{Data sets and the method}

\subsection{Data}
\label{Sec:Data}

The used data-sets are:

\begin{itemize}
\item
a) {\it IntCal09} $\Delta^{14}$C global series: 11000 BC -- 1900 AD, 5-yr time resolution \citep{reimer_09}.

\item
b)  {\it SB93} $\Delta^{14}$C global annual series: 1511 -- 1900 AD \citep{stuiver93}.

\item
c) {\it Dye3} $^{10}$Be Greenland annual series: 1424--1985 AD \citep{mccracken04}.

\item
d) {\it NGRIP} $^{10}$Be Greenland annual series: 1389--1994 AD \citep{berggren09}.

\item
e) {\it SP} $^{10}$Be South Pole Antarctic series: 850--1950 AD, quasi-decadal \citep{raisbeck90,bard97}.

\item
f) {\it DF} $^{10}$Be Dome Fuji Antarctic series: 695--1880 AD, quasi-decadal \citep{horiuchi08}.

\item
g) {\it GRIP} $^{10}$Be Greenland series: 7380 BC--1640 AD, quasi-decadal \citep{yiou97,vonmoos06}.

\item
h) {\it M12} $\Delta^{14}$C Japanese annual series: 750 -- 820 AD, annual/biannual \citep{miyake12}.
\end{itemize}

\subsection{Model computations}
\label{Sec:model}

In order to evaluate possible SPE signatures in the data, we used model computations of
 the isotope production by energetic particles, assuming instant injection of SEP into the atmosphere
 and calculating the expected isotope response in terrestrial archives.

As the reference event, we considered an extreme SPE of 23-Feb-1956 (SPE56) \citep{meyer56} with a very
  hard spectrum \citep{tylka09,usoskin_ACP_11}.
While SPE56 was the strongest observed ground-level enhancement (GLE), $>4000$\% in the count rate of the Leeds NM,
 it had relatively modest fluence $F_{30}=10^9$ cm$^{-2}$ \citep{shea90}.
Yet sometimes SPEs with large fluence but soft spectrum occur, e.g.,
 a modest (only 10\% at the polar Oulu NM) GLE event of 04-Aug-1972 (SPE72) with high fluence $F_{30}=5\cdot 10^9$ cm$^{-2}$.
Since cosmogenic isotopes are produced by the most energetic part of the SEP energy spectra ($>1$ GeV),
 we consider here the SPE56 scenario.

Response of $^{10}$Be to SPEs was calculated similar to \citet{usoskin_GRL_SCR06}
 but using an updated $^{10}$Be yield function \citep{kovaltsov_Be10_10}, the corresponding geomagnetic model,
 and an intermediate atmospheric mixing model \citep[polar tropospheric and hemispherical stratospheric mixing -- see][]{field06,heikkila09}.
The calculated $^{10}$Be production is $7.5\cdot 10^4$ atoms/cm$^2$ for the SPE56 scenario.
Because of the $^{10}$Be stratospheric residence time \citep{heikkila09}, a possible SPE peak in $^{10}$Be data
 can be 2--3-yr long.
Since present models cannot convert the $^{10}$Be production into concentrations in ice,
 they are typically assumed to be directly proportional to each other \citep[e.g.,][]{mccracken04}.
For the normalization we use the reference period 1850--1900 AD with moderate solar activity:
 the galactic cosmic ray (GCR) modulation potential 443 MV \citep{alanko07} was close to
 the mean Holocene value \citep{usoskin_AA_07}.
Accordingly, the reference period $^{10}$Be production is
 $\langle Q\rangle=3.45\cdot 10^{-2}$ at/cm$^{2}$/sec, or $1.1\cdot 10^6$ at/cm$^2$/year.
This value is scaled to the mean measured concentration for the reference period, in each ice core
 (1.025, 1.86, 3.47 and 2.76 $10^4$ at/g for {\it Dye3, NGRIP, SP} and {\it DF} series, respectively).
The {\it GRIP} series was normalized to the entire interval.

\begin{figure}
\begin{center}
\resizebox{12cm}{!}{\includegraphics{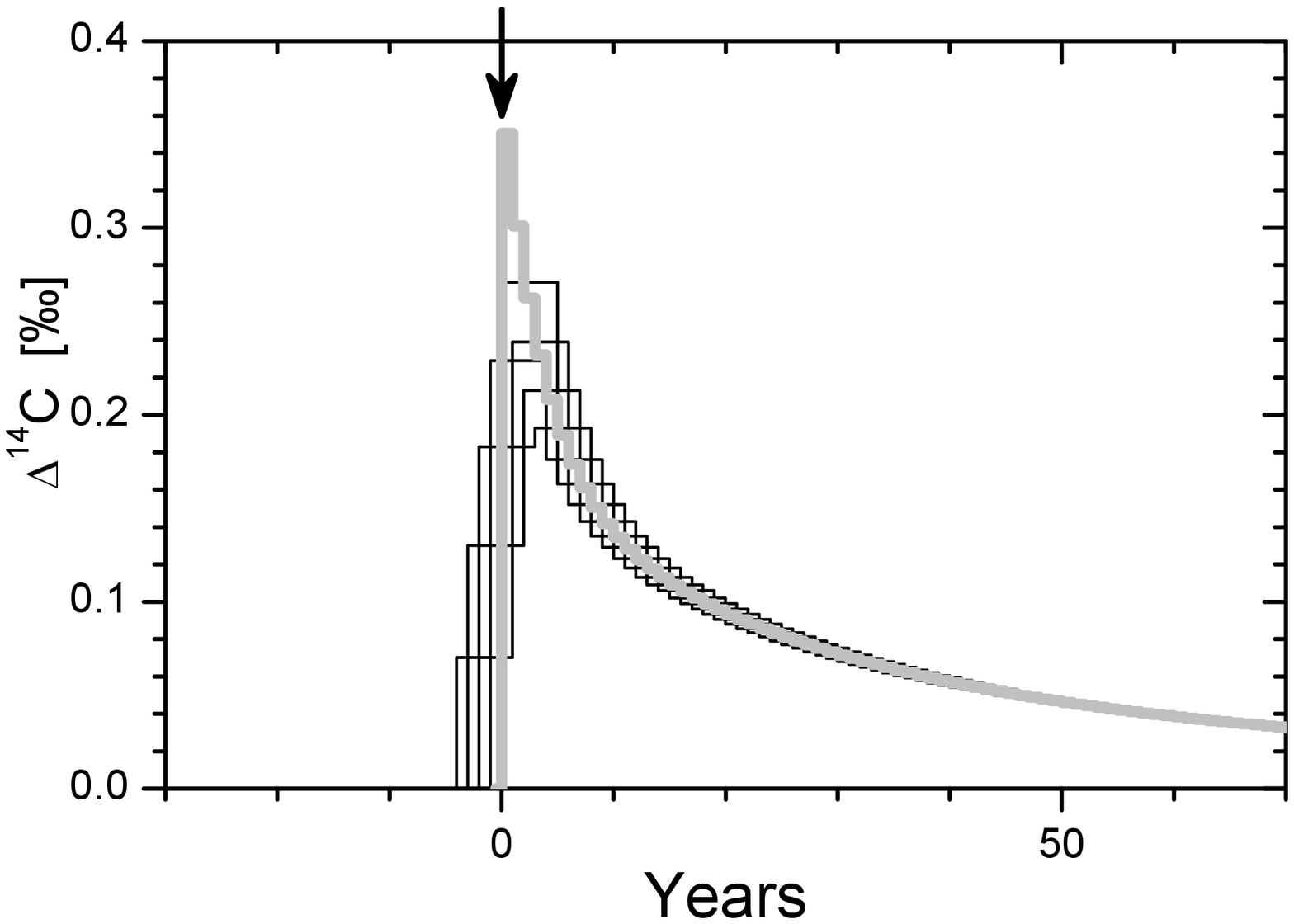}}
\end{center}
\caption{
Response of the relative tropospheric concentration $\Delta^{14}$C to SEP56 ($t=0$ shown by the arrow),
 with annual (grey curve) and 5-yr (black) time resolutions, depending on the 5-yr bins boundaries.
\label{Fig:14C_event}}
\end{figure}

The expected radiocarbon response of $\Delta^{14}$C was calculated in two steps.
First, the global production $Q$ was computed using a new production model
 \citep{kov12}, yielding $2.9\cdot 10^6$ at/cm$^2$ for SPE56.
Next, the instant atmospheric injection of $^{14}$C ($Q$) was
 passed through a 5-box carbon cycle model \citep{damon04}, and the tropospheric $\Delta^{14}$C was calculated.
Fig.~\ref{Fig:14C_event} shows the expected response in $\Delta^{14}$C for SPE56, which
 is extended over decades and is characterized by a sharp increase and exponential decay, with the peak's FWHM 15--20 years.
Accordingly, we look for a signature like that in the $\Delta^{14}$C data.
The SPE56 peak in the annual (5-year) $\Delta^{14}$C data is 0.35 ($0.19\div0.27$) permille,
 viz. below measurement errors of $\approx 2$ permille.
Therefore, an SPE needs to be a factor $X_{\rm SPE56}=10$ greater than SPE56 to be observable in the
 radiocarbon data.

An SPE72-type soft event would require a 40-fold greater $F_{30}$, with respect to SPE56, to produce the
 same amount of cosmogenic isotopes.
Accordingly, the estimates obtained with the reference SPE56 spectrum should be 40-fold enhanced to correspond to
 a soft-spectrum SPE72 scenario.

\section{Evaluating SEP fluence in the past}

First we analyze data series with the annual resolution (b--d in Section~\ref{Sec:Data}), looking
 for potential signatures of SPEs.

For the {\it NGRIP} series we looked for peaks with the magnitude $\ge 1.3\cdot 10^4$ at/g and duration $\le 3$ years.
Seven candidates were selected: 1436, 1460, 1650, 1719, 1810, 1816 and 1965 AD.
For the $^{10}$Be {\it Dye3} series five candidates (magnitude $\ge 0.6\cdot 10^4$ at/g, duration $\le 3$ years)
 were selected: 1462, 1479, 1505, 1512 and 1603 AD.
For the $^{14}$C {\it SB93} series no suitable candidates (peak with sharp rise and gradual decay) were found.
Since the found candidates do not coincide in time, we performed a cross-check.
For each candidate found in series $A$ we calculated, using the model (Sect.~\ref{Sec:model}),
 the corresponding isotope production $Q$ and the expected response in series $B$, and checked if it is
 consistent with the data within $\pm 2$ years, and vise versa.
For example, a 1436 AD peak in the {\it NGRIP} series must be accompanied by a strong peak in the {\it Dye3} series, which was not observed.
Also, a {\it NGRIP} 1965 AD peak is rejected by direct cosmic ray observations.
Using this cross-check seven candidates can be excluded: 1436, 1650 1816 and 1965 AD in
 {\it NGRIP} and 1479, 1512 and 1603 AD in the {\it Dye3} series.
The {\it SB93} series is insensitive to check the $^{10}$Be-based candidates.
Finally, five peaks pass through the check as potential SPE signatures (first block in Table~\ref{Tab:SPE_1}).
\begin{table}
\caption{identified SPE candidates: approximate year, dataset, and the $F_{30}$ fluence [cm$^{-2}$]
 evaluated for the SPE56 scenario. Scaling to SPE56 is given in parentheses.}
\begin{tabular}{ccl}
\hline
SPE year & Series & $F_{30}$ ($X_{\rm SPE56}$)\\
\hline
1460--1462 AD$^\dagger$ & NGRIP(1460) & $1.5\cdot 10^{10}$ (15) \\
              & Dye3 (1462) & $9.7\cdot 10^{9}$ (10) \\
1505 AD       & Dye3        & $1.3\cdot 10^{10}$ (13) \\
1719 AD &     NGRIP  & $1\cdot 10^{10}$ (10)  \\
1810 AD &     NGRIP  & $1\cdot 10^{10}$ (10) \\
\hline
 8910 BC & IntCal09 & $2.0\cdot 10^{10}$ (20) \\
 8155 BC & IntCal09 & $1.3\cdot 10^{10}$ (13) \\
 8085 BC & IntCal09 & $1.5\cdot 10^{10}$ (15) \\
 7930 BC & IntCal09 & $1.3\cdot 10^{10}$ (13) \\
 7570 BC & IntCal09 & $2.0\cdot 10^{10}$ (20) \\
 7455 BC & IntCal09 & $1.5\cdot 10^{10}$ (15) \\
 6940 BC & IntCal09 & $1.1\cdot 10^{10}$ (11) \\
 6585 BC & IntCal09 & $1.7\cdot 10^{10}$ (17) \\
 5835 BC & IntCal09 & $1.5\cdot 10^{10}$ (15) \\
 5165 BC & GRIP & $2.4\cdot 10^{10}$ (24) \\
 4680 BC & IntCal09 & $1.6\cdot 10^{10}$ (16) \\
 3260 BC & IntCal09 & $2.4\cdot 10^{10}$ (24) \\
 2615 BC & IntCal09 & $1.2\cdot 10^{10}$ (12) \\
 2225 BC & IntCal09 & $1.2\cdot 10^{10}$ (12) \\
 1485 BC & IntCal09 & $2.0\cdot 10^{10}$ (20) \\
 95 AD & GRIP & $2.6\cdot 10^{10}$ (26) \\
 265 AD & IntCal09 & $2.0\cdot 10^{10}$ (20)\\
780 AD$^\dagger$ & IntCal09 & $2.4\cdot 10^{10}$ (24) \\
                   & M12 & $4\cdot 10^{10}$ (40)$^*$\\
                   & DF & $4.5\cdot 10^{10}$ (45)$^*$ \\
 1455 AD$^\dagger$ & SP & $7\cdot 10^{10}$ (70)$^*$ \\
\hline
\end{tabular}
\label{Tab:SPE_1}
\\$^\dagger$ Discussed separately.\\
$^*$ Overestimate.
\end{table}
Only one candidate (1460--1462 AD) is present in both series (discussed later).

Next we considered data with rougher time resolution.
In each of the $^{10}$Be data series (e--g in Section~\ref{Sec:Data}) with quasi-decadal resolution
 we searched for distinctive single peaks.
Two candidates were identified in the {\it GRIP} series: ca. 5165 BC and 95 AD.
One candidate was identified in the {\it SP} series ca. 1455 AD, probably related
 to the ca. 1460 event in {\it NGRIP} and {\it Dye3} annual series.
Two candidates were identified in the {\it DF} series, ca. 780 AD and 1805 AD.
A search for signatures (sharp $\le 10$-yr increase, FWHM 15-30 years, magnitude $>2$ permille)
 in the {\it IntCal09} $^{14}$C series yields many candidates (see an example in Fig.~\ref{Fig:AD780}A).
We also made a cross-check of all the candidates between different series.
For instance, the 1805 AD peak in the {\it DF} series is rejected basing on the annual {\it NGRIP} and {\it Dye3} data.
About half of the candidates in the IntCal09 series after 7000 BC are rejected basing on the $^{10}$Be data.
Therefore, half of the identified candidates before 7000 BC can also be spurious.

Finally, we list the candidates passing the cross-check, in Table~\ref{Tab:SPE_1}.
One can see that a 10-fold SPE56 $X_{\rm SPE56}=10$ is the detection limit in cosmogenic isotope data.
Only two candidates appear in more than one series - ca. 780 AD and 1460 AD, which are discussed below in detail.

\subsection{Event of ca. 1460 AD}

\begin{figure}
\begin{center}
\resizebox{12cm}{!}{\includegraphics{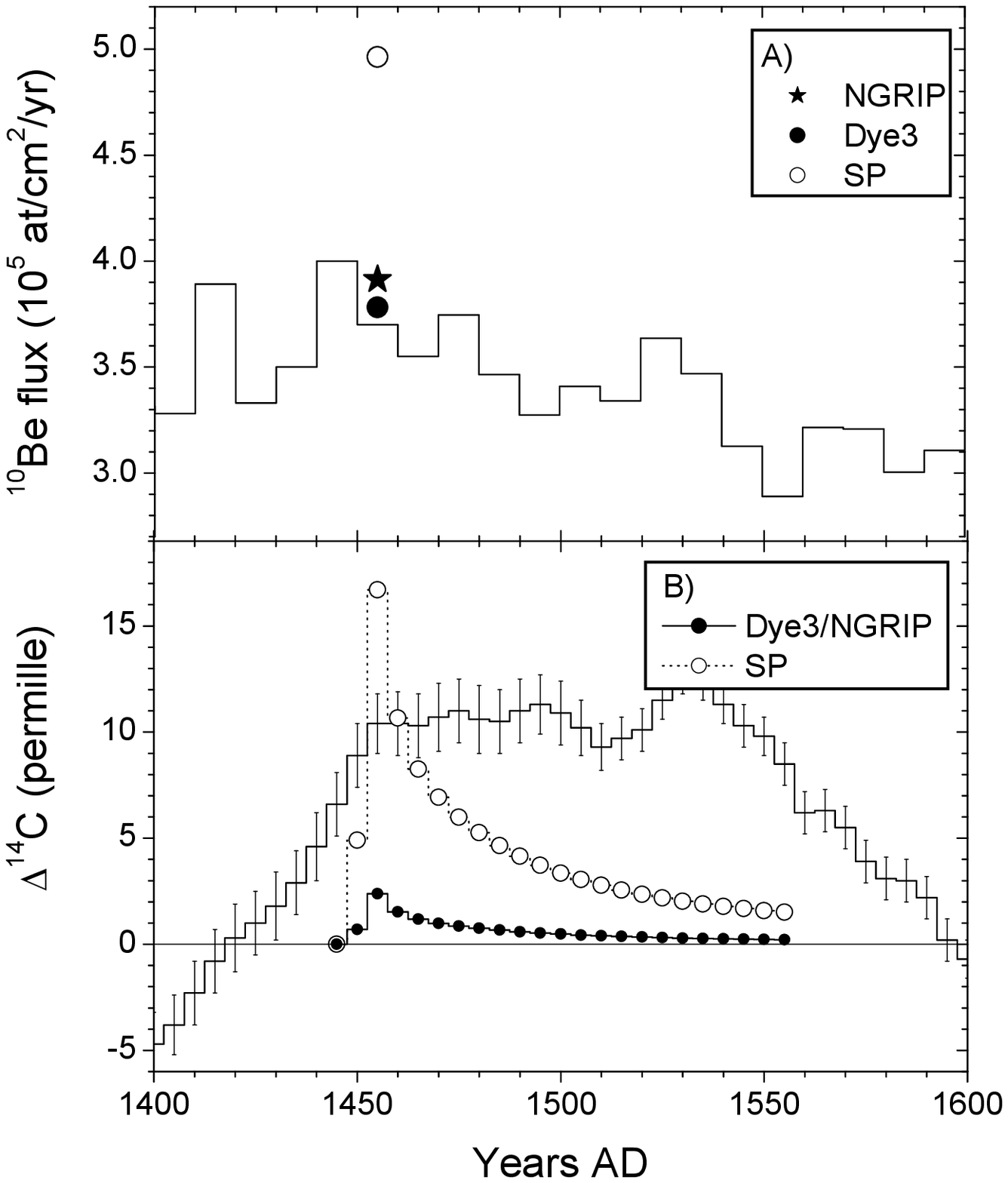}}
\end{center}
\caption{Test of a possible SPE ca. 1460 AD.
A) {\it DF} $^{10}$Be series, with points corresponding to the SEP-production estimated from {\it NGRIP, Dye3} and {\it SP} data series (see Legend).
B) {\it IntCal09} $\Delta^{14}$C series along with the expected signal estimated from {\it Dye3/NGRIP} and {\it SP} data series.
\label{Fig:DF1460}}
\end{figure}

Several analyzed series show a significant peak around 1460 AD (Table\ref{Tab:SPE_1}).
This was noticed earlier and ascribed to a very strong SPE or to a supernova explosion \citep[e.g.][]{berggrenPhD,mccracken04,delaygue11}.
We assume here that it is a SPE signature and evaluate its parameters.
Annual {\it NGRIP} and {\it Dye3} series depict distinct peaks in 1460 and 1462 AD, respectively, which roughly
 agree with each other, and require  $F_{30}\approx 10^{10}$ protons/cm$^2$
 for the SPE56 scenario.
Rougher resolved {\it SP} series depicts a very strong peak ca. 1455 AD, requiring the fluence 7 times greater than implied from the {\it NGRIP/Dye3} data.
Neither of other series ({\it GRIP, DF} or {\it IntCal09}) show peaks around that date allowing to evaluate the upper limit.
Fig.~\ref{Fig:DF1460}A shows a cross-check of this event vs. the {\it DF} series.
While the event strength evaluated from {\it NGRIP} and {\it Dye3} series is consistent with no
 clear signal in the {\it DF} data, the huge fluence implied from the {\it SP} data contradicts the {\it DF} data \citep[cf.][]{delaygue11}.
A similar analysis of the {\it GRIP} data also suggests that the {\it SP} peak is too high.
Fig.~\ref{Fig:DF1460}B shows the {\it IntCal09} data along with signals expected from {\it NGRIP/Dye3}
 and {\it SP} data, respectively, for the ca. 1460 AD candidate.
Because of the steep gradient in $\Delta^{14}$C in 1430--1470, it is impossible
 to distinguish a small 2.4 permille peak implied by the {\it Dye3/NGRIP} data, but the high signal implied by
 the {\it SP} peak apparently contradicts the data.
Thus, while all the analyzed datasets are consistent with a hypothesis of a strong SPE ca. 1460 AD,
 its fluence is likely about $10^{10}$ protons/cm$^2$ for the SPE56 scenario.
The very high fluence implied by the {\it SP} data contradicts other datasets.

\subsection{Event of ca. 780 AD}

\begin{figure}
\begin{center}
\resizebox{\hsize}{!}{\includegraphics{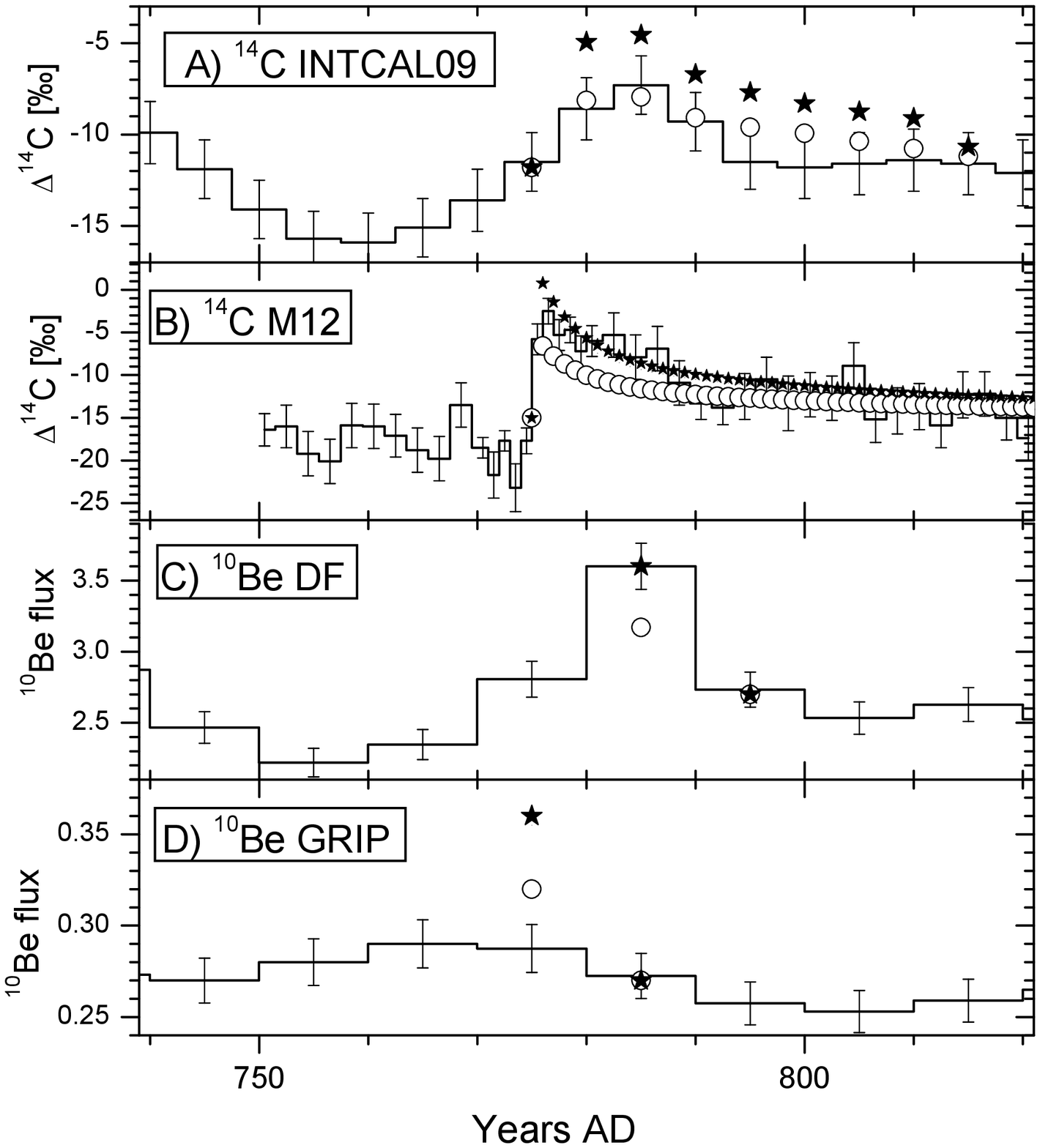}}
\end{center}
\caption{Test of a possible SPE ca. 780 AD.
Open dots and black stars correspond to the modelled response to a SEP event equal to $X_{SPE56}=24$ and $X_{SPE56}=45$, respectively.
A) {\it IntCal09} $\Delta^{14}$C data series.
B) {\it M12} $\Delta^{14}$C data series.
C) {\it DF} $^{10}$Be data series.
D) {\it GRIP} $^{10}$Be data series.
\label{Fig:AD780}}
\end{figure}

Another interesting candidate for a strong SPE is ca. 780 AD as manifested through
 distinct peaks in {\it DF}, {\it IntCal09} and {\it M12} data series (Fig.~\ref{Fig:AD780}).
According to these data-sets, the event was 25--50 times stronger than SPE56 with the fluence $F_{30}=(2\div 5)\cdot 10^{10}$ protons/cm$^2$.
The best fit for the {\it IntCal09} data (Fig.~\ref{Fig:AD780}A) is obtained for a $X_{SPE56}=24$ event started ca. 780 AD.
The {\it M12} data (Fig.~\ref{Fig:AD780}B) is better fit with a $X_{SPE56}=40$ event started in 775 AD,
 while the {\it DF} data (Fig.~\ref{Fig:AD780}C) is consistent with $X_{SPE56}=45$ event occurred between 780 and 790 AD.
However, the {\it GRIP} series (Fig.~\ref{Fig:AD780}D) depicts no peak at that period, which indicates that the fluence greater than $F_{30}=3\cdot 10^{10}$ protons/cm$^2$
 is inconsistent at 0.03 significance level.
Applying the cross-check, we found that the fluence implied by the {\it IntCal09} data
 ($X_{SPE56}=24$ -- see Table~\ref{Tab:SPE_1}) is consistent with the {\it GRIP} data, while both $X_{SPE56}=40$ and 45
 based on the {\it DF} and {\it M12} data, respectively, yield a too high peak in the {\it GRIP} series.
Accordingly, we conclude that the event of 780 AD is a possible candidate for a strong SPE with a consensus
 value $F_{30}\approx 3\cdot 10^{10}$ protons/cm$^2$.

\subsection{Carrington event of 1859 AD}

\begin{figure}
\begin{center}
\resizebox{12cm}{!}{\includegraphics{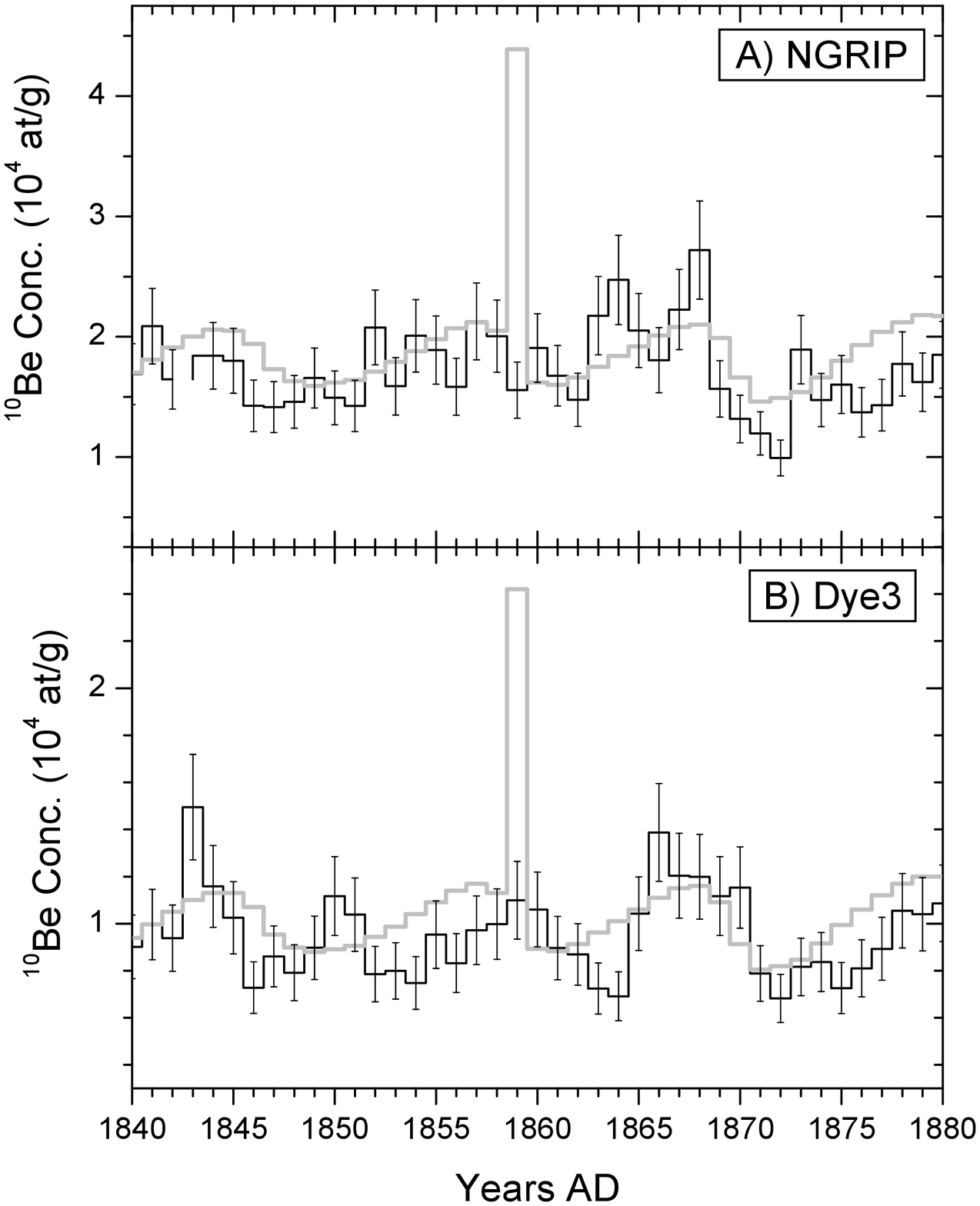}}
\end{center}
\caption{Testing a possible SEP event ca. 1859 AD.
The gray line depicts a model signal by GCR plus an addition of the annual
 production due to the Carrignton 1859 AD event according to \citet{mccracken01}.
A) Time series of the annual {\it NGRIP} $^{10}$Be data, with 15\% error bars.
B) Time series of the annual {\it Dye3} $^{10}$Be data, with 15\% error bars.
\label{Fig:1859}}
\end{figure}

The Carrington event of 1859 AD is often considered as an extreme SPE, with
 $F_{30}=1.8\cdot 10^{10}$ cm$^{-2}$ as estimated from the nitrate record in the Greenland Summit core \citep{dreschhoff98,mccracken01,shea06}.
Although none of the series analyzed here depicts a peak around 1859, we check if
 this proposed SPE is consistent with the cosmogenic data.
Using the GCR modulation reconstruction \citep{alanko07} and the $^{10}$Be production model \citep{kovaltsov_Be10_10},
 we calculated the expected $^{10}$Be concentration along with the additional production for 1859 AD, applying SEP parameters
 for the Carrington SPE according to Section 8 of \citet{mccracken01}.
The model computation is then confronted with the annual data of two Greenland series, {\it NGRIP} and
 {\it Dye3}, in Fig.~\ref{Fig:1859}.
The $^{10}$Be peak, expected from the nitrate data, is too strong, contradicting
 to the observed data in two Greenland sites.
Radiocarbon data and decadal $^{10}$Be series cannot resolve the Carrington peak.
Therefore, we conclude that the cosmogenic data do not support the hypothesis of
 a very strong SPE related to the Carrington flare.
%

\section{Discussion}

We propose, by analyzing data from different cosmogenic isotope records and performing
 their cross-check, a list of candidates for strong SPEs in the past (Table~\ref{Tab:SPE_1}).
The list is somewhat uncertain, as some events may be missing or some
 spurious peaks falsely identified.
However, the result is quite consistent in a statistical sense.

Four candidates with the $X_{\rm SPE56}$ factor 10$\div$15 are identified in the annual
 $^{10}$Be series {\it NGRIP} and {\it Dye3} for the last 600 years (first block of Table~\ref{Tab:SPE_1}).
Events with $X_{\rm SPE56}<10$ cannot be reliably identified.
We can also securely say that events with $X_{\rm SPE56}>20$ are not observed during that period (cf. Fig.~\ref{Fig:1859}).
In particular, a strong SPE related to the Carrington flare in 1859 contradicts these data.

We evaluate the probability $p$ of occurrence of such events, applying the Poisson distribution (assuming
 that SPEs are mutually independent).
For example, for four events observed during 600 years, the probability is
 $p=0.0077 _{-0.0045}^{+0.0073}$ year$^{-1}$ (90\% confidence interval), which is higher
 than the naively taken $4/600=0.0067$ year$^{-1}$.
If no event with $F_{30}>2\cdot 10^{10}$ cm$^{-2}$ is observed over 600 years, the corresponding
 one-sided 90\% confidence interval is $p=0\div0.0027$ year$^{-1}$, with the median probability
 0.0012 year$^{-1}$ (once per 850 years).

The last 4--6 centuries cover the full range of solar activity, from Grand minima
 to the maximum \citep{usoskin_LR_08} serving as an archetype of the solar variability.
Therefore we expect that this result is consistent with longer scales.
On the multi-millennial scale (the Holocene -- 11 millennia), 19 candidates are identified
 (second block in Table~\ref{Tab:SPE_1}) with $X_{\rm SPE56}=10\div30$.
A few candidates with $X_{\rm SPE56}>50$ were rejected by the cross-check.
This gives the average occurrence rate of such SPEs roughly two per millennium.
Moreover, we can securely say that during the Holocene there was no events
 with $X_{\rm SPE56}>50\div100$, placing a strong upper limit on the SPE strength.
We found no apparent relation of the SPE candidates occurrence and the solar activity level \citep{usoskin_AA_07}.

\begin{figure}
\begin{center}
\resizebox{\hsize}{!}{\includegraphics{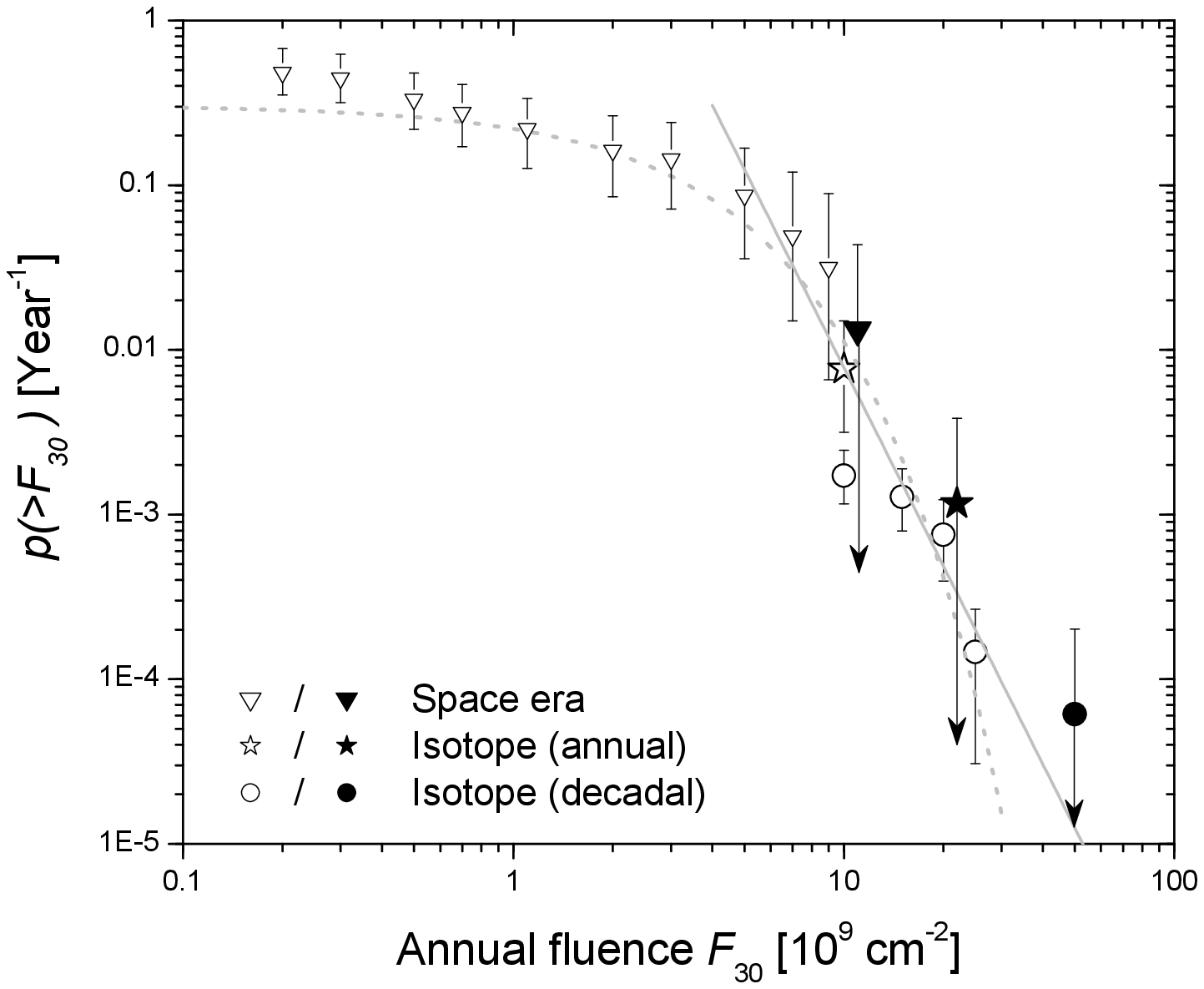}}
\end{center}
\caption{Probability of occurrence of the annual fluence ($>30$ MeV) exceeding the given value $F_{30}$,
 as evaluated from the data for the space era 1956--2008 (triangles),
 cosmogenic isotope annual data (stars) and cosmogenic isotope decadal data (circles).
Open symbols correspond to the statistics of ``observed'' data, while filled symbols denote the
 conservative upper limit (no events greater than X observed during Y years).
Error bars correspond to the 90\% confidence level evaluated assuming the Poisson statistics.
Grey dotted and solid lines depict the best-fit exponent and power law approximations
 for $F_{30}>7\cdot 10^9$ protons/cm$^2$.
\label{Fig:D_all}}
\end{figure}

We summarize our findings in a plot (Fig.~\ref{Fig:D_all}) of the integral probability of strong SPE occurrence.
X- and Y-axis give the $F_{30}$ fluence and the probability
 of occurrence of SPE with the $>30$ MeV proton annual fluence exceeding $F_{30}$, for different datasets.
The measured annual fluences for the space era (1956--2008) are shown by open triangles \citep[][Shea, private communication 2012]{shea90}.
Since the space era coincides with the unusually active Sun \citep{solanki_nat_04}, this probabilities may be
 higher than for the typical medium activity Sun.
The black triangle reflects the fact that no SPE with $F_{30}>10^{10}$ cm$^{-2}$ was observed during 53 years.
The open star corresponds to the four SPE candidates from the annual $^{10}$Be data (first block in Table~\ref{Tab:SPE_1}),
 while the filled star corresponds to no SPE with $F_{30}>2\cdot 10^{10}$ cm$^{-2}$ found during 600 years.
Open circles represent the candidates found in the rougher time series (second block in Table~\ref{Tab:SPE_1}),
 grouped into four points $F_{30}>1$, 1.5, 2 and 2.5 $10^{10}$ cm$^{-2}$, respectively.
Since the lower fluence value is probably underestimated because of the detection threshold,
 we ignore the first open circle, using instead the star-symbolled point from the annual $^{10}$Be datasets.
The filled circle corresponds to no event with $F_{30}>5\cdot 10^{10}$ cm$^{-2}$ found over 11 millennia.
One can see that there is a break in the distribution at about $F_{30}=5\cdot 10^{9}$ cm$^{-2}$.
This break \citep[cf.][]{mccracken01,hudson10} may be related to a streaming limit \citep{reames99} and/or
 limitations in the energetics of solar flares/CME \citep{fletcher11}.
Our new results put a stronger constraint on the distribution.
For example, a SPE $X_{\rm SPE56}>20$ is expected only once per millennium, and
 no $X_{\rm SPE56}>50$ event is expected over tens of millennia.
We try to fit the tail of the distribution ($F_{30}>7\cdot 10^{9}$ cm$^{-2}$) by two
 simple models: exponential ($p\propto \exp{(-0.33\cdot10^{-9}\cdot F_{30})}$) and power law
  ($p\propto F_{30}^{-4}$) -- see grey lines in the Figure.
The data does not allow us to distinguish between the two shapes.

We also compare our results with the average SEP flux over time scales of
 thousands--millions years, estimated from lunar rocks, which ranges 21$\div$56 (cm$^2$ sec)$^{-1}$
 \cite[e.g.][]{reedy99,nishiizumi09}.
Time averaging of the results from Fig.~\ref{Fig:D_all} using the above fits for the
 tail yields the average flux $\approx 38$ (cm$^2$ sec)$^{-1}$, which is composed of about
 35 (cm$^2$ sec)$^{-1}$ for the space era record \citep[e.g.,][]{shea02,reedy12} with an addition of
 3.2 (cm$^2$ sec)$^{-1}$ due to the distribution tail based on cosmogenic isotope data (both exponent and
 power law give similar results), suggesting that extreme SPEs contribute only about 10\% to the total SEP fluence.
This is totally consistent with the assessments based on lunar rocks, giving an independent support to the
 present results.
For a SPE with a softer energy spectrum, like SPE72, the tail addition would be 40-times greater (Section~\ref{Sec:model})
 leading to the average SEP flux $\approx$150 (cm$^2$ sec)$^{-1}$, in contradiction with the lunar rock data.
Therefore, extreme SPEs found in the cosmogenic isotope records must have (on average) hard energy spectra.

\section{Conclusions}

We have evaluated the probability of occurrence of extreme SPEs based on data of cosmogenic
 isotopes $^{14}$C and $^{10}$Be in terrestrial archives, spanning over the time scale from centuries
 to 11 millennia (Fig.\ref{Fig:D_all}).
We identified four potential candidates for SPEs with $F_{30}=(1\div 1.5)\cdot 10^{10}$ cm$^{-2}$, and show that
 no event with $F_{30}>2\cdot 10^{10}$ cm$^{-2}$ existed, over the last 600 years using annually resolved $^{10}$Be data.
In particular, the extreme Carrington SPE of 1859 AD contradicts these data.
From rougher resolved data we identified 19 SPE candidates (Table~\ref{Tab:SPE_1}) with $F_{30}=(1\div 3)\cdot 10^{10}$ cm$^{-2}$,
 and clearly no event with $F_{30}>5\cdot 10^{10}$ cm$^{-2}$, over the last 11400 years.
Two events appear in different series, ca. 780 AD and 1460 AD making them strong candidates to extreme SPEs.
This gives a new strict observational constraint on the occurrence probability of extreme SPEs.
Practical limits can be set as $F_{30}\approx 1$, $2\div 3$ and 5 $10^{10}$ cm$^{-2}$ (10-, 20$\div$30- and 50-times greater than SPE56),
 for the occurrence probability of $10^{-2}$, $10^{-3}$ and $10^{-4}$ year$^{-1}$, respectively.
The mean SEP flux is found as $\approx 40$ (cm$^2$ sec)$^{-1}$ in agreement with estimates from the lunar rocks.
On average, extreme SPEs contribute about 10\% to the total SEP fluence.

We note that the present result tends to represent an upper limit for the SPE occurrence, since we
 explicitly assume that every peak in one data series, consistent with other series, is a signature of SPE.
This may be not exactly correct as some of the peaks may be spurious.
Accordingly, our results should be interpreted as a conservative upper limit of the SPE occurrence
 near Earth.
Application of the results to the Sun is not straightforward, especially for most energetic events
 with low statistics, because the propagation of SEP in the interplanetary space may greatly affect
 geo-efficiency of SPE, and accordingly, their ability to become detectable in the cosmogenic isotope
 data series.
Given these uncertainties, the present results should be considered with a precision of up to an order of magnitude. 

\acknowledgments
{We thank M.A. Shea for the data on observed annual SEP fluences.
GAK was partly supported by the Program No.22 of the Presidium RAS, and acknowledges also support from the Academy of Finland
 and Suomalainen Tiedeakatemia (V\"ais\"al\"a foundation).
}



\end{document}